\documentclass[twocolumn,twocolappendix]{aastex631}
\usepackage{amsmath,amssymb}
\usepackage{multirow}
\usepackage{lipsum}
\usepackage{float}

\begin{document}

\title{New Insights into Type-I Solar Noise Storms from High Angular Resolution Spectroscopic Imaging with the upgraded Giant Metrewave Radio Telescope}

\author[0000-0002-2325-5298]{Surajit Mondal}
\affiliation{Center for Solar-Terrestrial Research, New Jersey Institute of Technology, 323 M L King Jr Boulevard, Newark, NJ 07102-1982, USA}

\author[0000-0001-8801-9635]{Devojyoti Kansabanik}
\affiliation{Cooperative Programs for the Advancement of Earth System Science, University Corporation for Atmospheric Research, 3090 Center Green Dr, Boulder, CO, USA 80301}
\affiliation{NASA Jack Eddy fellow hosted at the Johns Hopkins University Applied Physics Laboratory, 11001 Johns Hopkins Rd, Laurel, USA 20723}

\author[0000-0002-4768-9058]{Divya Oberoi}
\affiliation{National Centre for Radio Astrophysics, Tata Institute of Fundamental Research, S. P. Pune University Campus, Pune 411007, India}

\author[0009-0006-3517-2031]{Soham Dey}
\affiliation{National Centre for Radio Astrophysics, Tata Institute of Fundamental Research, S. P. Pune University Campus, Pune 411007, India}

\correspondingauthor{Surajit Mondal}
\email{surajit.mondal@njit.edu, surajit4444mondal@gmail.com }

\begin{abstract}
Type-I solar noise storms are perhaps the most commonly observed active radio emissions from the Sun at meter-wavelengths. Noise storms have a long-lived and wideband continuum background with superposed islands of much brighter narrowband and short-lived emissions, known as type-I bursts. There is a serious paucity of studies focusing on the morphology of these two types of emissions, primarily because of the belief that coronal scattering will always wash out any features at small angular scales. However, it is important to investigate their spatial structures in detail to make a spatio-temporal connection with observations at extreme-ultraviolet/ X-ray bands to understand the detailed nature of these emissions.
In this work, we use high angular resolution observations from the upgraded Giant Metrewave Radio Telescope to demonstrate that it is possible to detect structures with angular scales as small as $\sim 9\arcsec$, about three times smaller than the smallest structure reported to date from noise storms. Our observations also suggest while the individual type-I bursts are narrowband in nature, the bursts are probably caused by traveling disturbance(s) inducing magnetic reconnections at different coronal heights, and thus leading to correlated change in the morphology of the type-I bursts observed at a wide range of frequencies.
\end{abstract}


\section{Introduction}\label{sec:introduction} 
Solar type-I noise storms are the most common active emission features seen at meter-wavelength. These are characterized by short duration ($\sim 0.1–10$ s) intense narrowband ($\sim$ a few MHz) bursts (type-I bursts) superposed on a long duration wideband ($\gtrsim 100$ MHz) continuum emission \citep{mclean1967}. Noise storms can last up to several days \citep{elgaroy1977, suresh2017}. The emission often has very strong circular polarization \citep[e.g.][etc.]{zlobec1971, ramesh2011, mugundhan2018,mccauley2019} and is thought to arise due to plasma emission from nonthermal accelerated electrons trapped in coronal loops \citep{sakurai1971, melrose1980}. Imaging studies of type-I noise storms have revealed that these are generally associated with active regions and are often compact in size \citep{mugundhan2018, mccauley2019, mohan2019b}. However, most of the imaging studies have been done using instruments either having poor spatial resolution ($\gtrsim 1\arcmin$) or using disk-integrated dynamic spectra, which cannot provide any spatial information. This makes it harder to associate type-I sources with the sources seen using higher angular resolution extreme ultraviolet (EUV)/X-ray observations. However, doing so is crucial, not only for understanding type-I noise storms in greater detail, but also for investigating the suggested relationship between noise storms, and solar flares and coronal mass ejections \citep{Vourlidas2020}.

There were a few studies that used data at high spatial resolution ($\lesssim 10\arcsec$) observations to study type-I noise storms. However barring \citet{mugundhan2018}, none of the earlier studies detected structures with an angular scale smaller than $\approx 30\arcsec-40\arcsec$ \citep{lang1987, zlobec1992, kerdraon1988, mercier2015,mercier2006}. However, it is well known that structures much smaller than that are visible in the EUV wavelengths. Based on the understanding that local electron acceleration is powering the noise storm, it is natural to expect structures at similar scales in noise storms as well. It is not immediately clear why observations providing an angular resolution of $3\arcsec$ can miss all structures smaller than $40\arcsec$ in multiple instances. These non-detections of small scale spatial structures led to the idea that scattering due to coronal density inhomogeneities smooths out these smaller angular scales \citep[][]{bastian1994, kontar2019} at radio wavelengths. 
The exact limiting angular scale below which all spatial structures get smeared out due to scattering also depends on the assumptions made regarding the turbulent nature of the solar corona and can range from sub-arcsecond scales to about an arcminute \citep{subramanian2011}. Hence, it is  crucial to perform spectroscopic snapshot imaging studies of type-I noise storms using new instruments capable of providing high angular observations. This will not only be able test this paradigm, but also has the potential to place stronger constraints on solar wind turbulence, while also improving our understanding regarding type-I bursts and noise storms.

With this objective, we observed the Sun using the upgraded Giant Metrewave Radio Telescope \citep[uGMRT,][]{gupta2017}. We not only probe the details of type-I noise storms at high spatial resolution but also investigate how their morphology change with time and frequency. Section \ref{sec:cur_stat} discusses the current status of high angular resolution observations of type-I noise storms. The observation and calibration details are provided in Section \ref{sec:obs_cal}. The results obtained from this analysis are provided in Section \ref{sec:results}. In Section \ref{sec:discussion}, we place our results in the overall context of noise storms and the impact of scattering, and finally in Section \ref{sec:conclusion} we present the conclusions from this investigation.

\section{Current Status of High Angular Resolution Imaging of type-I Noise Storms}\label{sec:cur_stat}

\citet{lang1987} observed type-I noise storm using the VLA at 328 MHz band, with a spatial resolution of $9\arcsec \times 9\arcsec$, for around five hours duration. They found that the noise storm consisted of multiple sources, with sizes of about $40\arcsec$, located within an extended source. \citet{zlobec1992} also did similar observations but did not detect any power on baselines longer than 5000$\lambda$  (which corresponds to about $40\arcsec$), where $\lambda$ is the observing wavelength. Hence they concluded that the smallest morphological structure inside noise storms at 333 MHz band cannot be smaller than $40\arcsec$. \citet{mercier2006} and \citet{mercier2015} combined data from the legacy Giant Metrewave Radio Telescope \citep[GMRT,][]{Swarup2000} and NRH, and produced snapshot images of noise storm with an angular resolution of approximately $20\arcsec$ at 236--327 MHz. The minimum size of the structures they detected ranges from 31$\arcsec$--35$\arcsec$.
Using these high resolution images, they concluded that the noise storm comprised of several knots, surrounded by a more extended emission. 
Using a two-element interferometer of baseline length $\sim200$ km, \citet{mugundhan2018} detected source sizes $\leq 15\arcsec$ at 53 MHz. 
It is generally well accepted that scattering smears out the solar radio emission and prevents the detection of small-scale structures \citep{bastian1994,kontar2019}
Observations by \citet{mugundhan2018}, however, suggest that, at least in some instances, small source sizes can indeed be observed. 
There are also suggestions that the lack of evidence in favor of smaller source sizes can simply be a consequence of angular resolution of instruments typically used for solar observations \citep{subramanian2011}.In this study, we use high angular resolution observations from the uGMRT to examine this question.

\begin{figure*}
    \centering
    \includegraphics[trim={0.4cm 0.8cm 1cm 0.3cm},clip,scale=0.31]{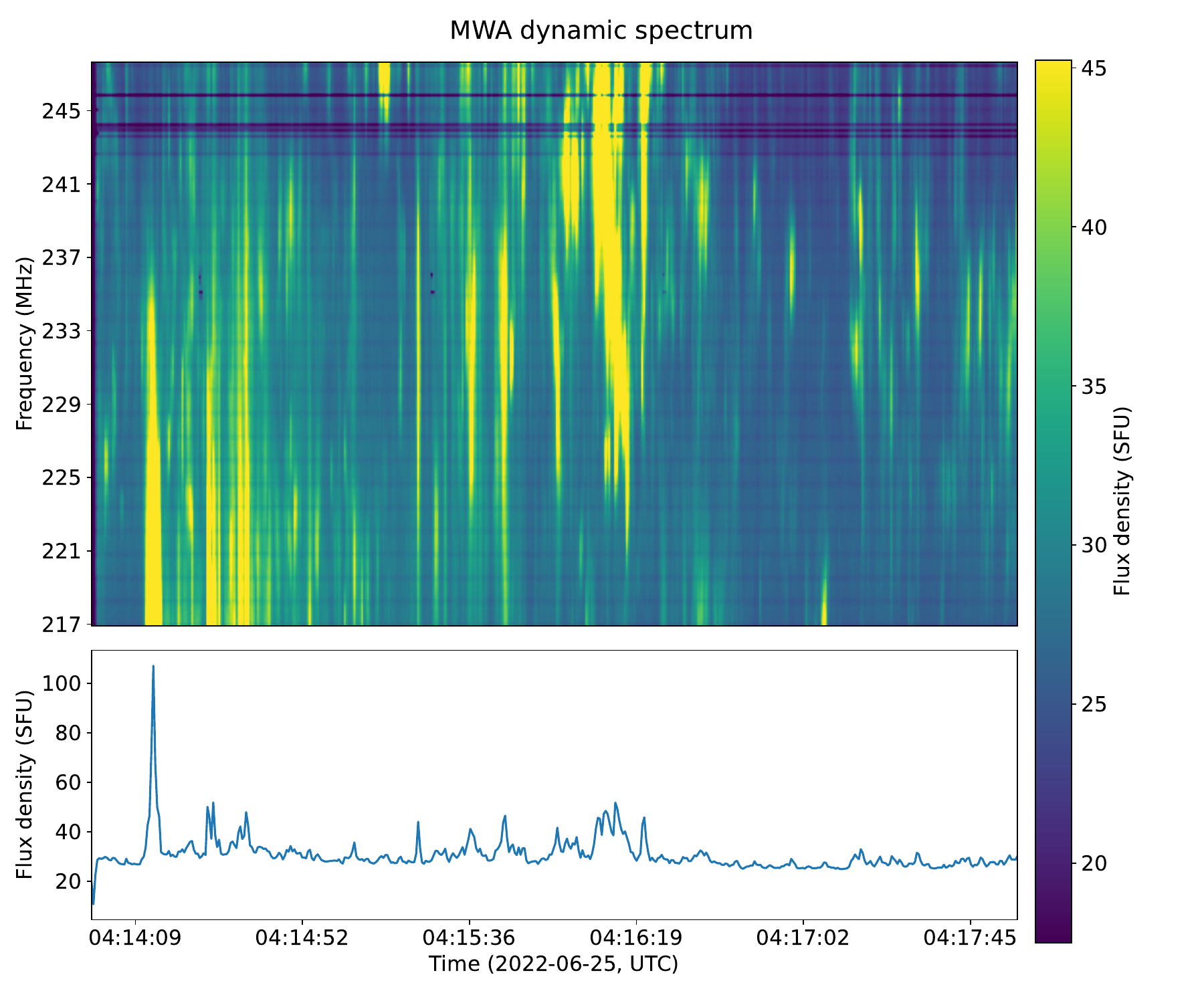}\includegraphics[trim={0.4cm 0.8cm 1cm 0.3cm},clip,scale=0.31]{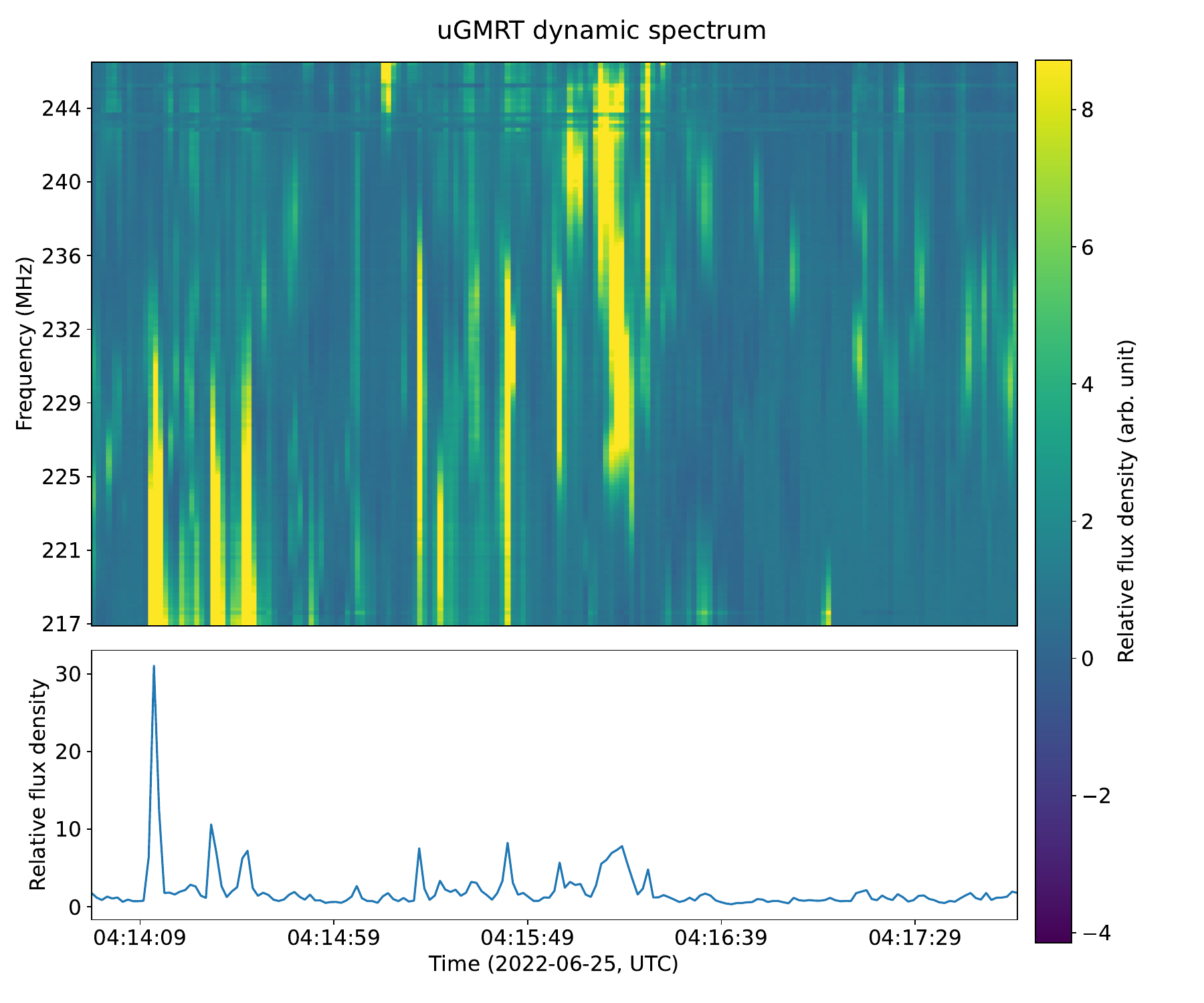}
     \caption{Left panel: Stokes I flux density calibrated dynamic spectrum of the MWA shown at the top and the time series of the band-averaged flux density is shown at the bottom. Right panel: Stokes I dynamic spectrum of uGMRT normalized using median bandshape is shown at the top and the time series of the band-averaged relative flux density is shown at the bottom. }
    \label{fig:mwads}
\end{figure*}

\section{Observation and Calibration}\label{sec:obs_cal}
The uGMRT is a radio interferometric array consisting of 30 antennas (each of them is 45 meter in diameter), spread over an area $\sim$25 km in diameter located at Khodad, India. The uGMRT provides an angular resolution of a few arcseconds over its entire observing frequency range. Simultaneous observations were also taken with the Murchison Widefield Array \citep[MWA,][]{Tingay2013}, at an overlapping frequency band. The MWA is located in the Western Australian desert. For this observation, the MWA was in its extended phase-II \citep{Wayth2018} configuration with 136 antenna tiles distributed over a footprint of $\sim$5 km. 

\subsection{uGMRT Observation Details}\label{subsec:obs}
We observe the Sun using the band-2 receivers of the uGMRT covering the frequency range 120--250 MHz. Sensitive radio telescopes are optimized to observe astronomical sources. Since the Sun has orders of magnitude higher flux density compared to the typical astronomical radio sources, one needs to attenuate the solar signal to maintain the entire signal chain of the instrument in the linear regime. Based on several tests done a-priori for developing standard observation procedure of solar observations using the uGMRT (Kansabanik et al., in prep), we used an attenuation of 30 dB for observing in uGMRT band-2.  Here, we present observations on 25 June 2022, recorded using the GMRT Wideband Backend \citep{reddy2017_gwb} between 03:30 to 04:30 UTC
\footnote{GMRT data available from \href{https://naps.ncra.tifr.res.in/goa/data/search}{GMRT Online Archive} is in International Atomic Time (TAI). We have converted GMRT timestamps to UTC in this article}.
The frequency and time resolution of these interferometric observations are 195.3125 kHz and 1.3 seconds, respectively. While full polarimetric data were recorded, a detailed polarimetric study is beyond the scope of this work and will be presented in a future paper (Dey et al., in prep). For calibrating the instrumental bandpass and absolute flux-density scale, we chose a bright flux-density calibrator, 3C 48 having a flux density of $\sim$ 42 Jy \citep{Perley_2017} at the uGMRT band-2. The flux density of 3C 48 is sufficiently large that it can be observed with the same attenuation as used for the solar signal. Since the total duration of observation was only 1 hour, no phase calibrator was observed to reduce observing overheads. The 1-hour duration is divided into four scans, each of 15-minutes duration. 

\subsection{Flagging and Calibration}\label{subsec:calflag}
We did the data analysis using the Common Astronomy Software Applications \citep[CASA,][]{CASA2022}. First, we performed basic flagging of bad antennas and radio frequency interference (RFI) affected spectral channels on 3C 48. This basic flagging was applied to both 3C 48 and solar scans. Next, we performed an automated RFI-flagging on the uncalibrated 3C 48 scan using \textsf{tfcrop} \citep{CASA2022} algorithm in \textsf{flagdata} task. Since the short baselines are more affected by correlated RFI, we performed automated flagging on the baselines shorter and longer than 1 km, separately. Solar emission can show large spectro-temporal variations, making it hard for an automated RFI-flagging algorithm to distinguish between RFI and genuine solar emission features. Hence, we did not perform automated RFI-flagging on solar scans.
\begin{figure*}
    \centering
    \includegraphics[trim={0cm 0cm 0cm 0cm},clip,scale=0.5]{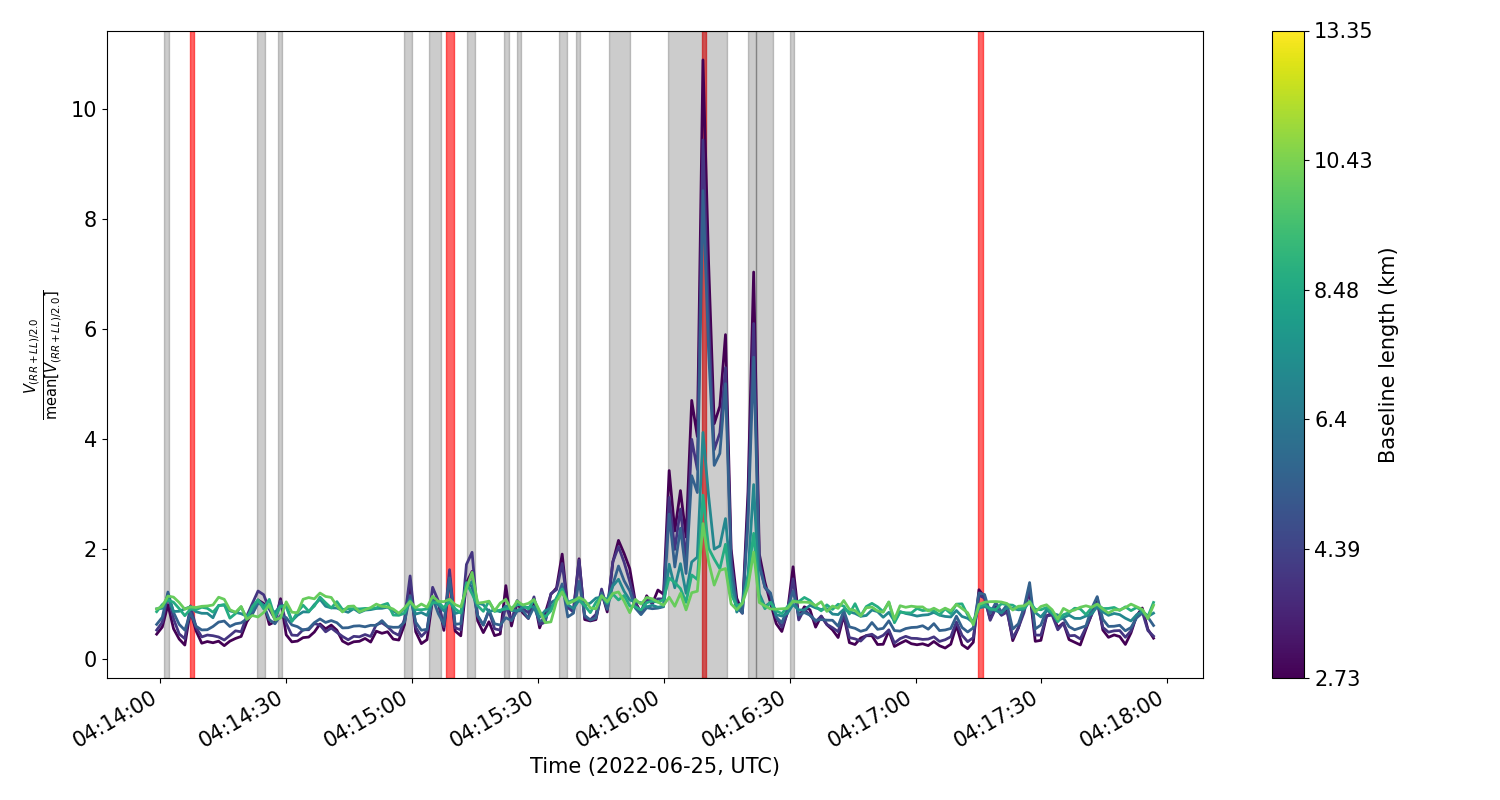}
    \caption{Timeseries of the amplitude of visibilities averaged over the two polarizations, RR and LL over the band 244-251 MHz. Each line represents the timeseries averaged over 50 consecutive baselines  (arranged in the order of baseline length) and divided by the mean amplitude of the timeseries. Vertical bars mark the peaks imaged in this study. Red bars show the peaks for which images are shown in Figure \ref{fig:no_enhancement_spw_0_35}.}
    \label{fig:lightcurve_0_35}
\end{figure*}

Once the first round of flagging was done, we determined the frequency-dependent complex gain of the instrument (instrumental bandshape) using the source model of 3C 48 \citep{Perley_2017} and \textsf{bandpass}. We then corrected 3C 48 observations for the estimated instrumental bandshape using \textsf{applycal} task. After applying the bandshape solutions, we performed an automated flagging on the residual data using \textsf{rflag} \citep{CASA2022} algorithm. This was followed by some manual inspection for identifying low-level RFI and flagging them as well. The presence of residual RFI can affect the determination of the instrumental bandshape. It was, hence, determined again after this round of flagging and then applied to the solar scans.

We then performed flagging on bandshape-corrected solar scans by manual inspection. Since solar emission varies across time and frequency, it is hard to identify the RFI in the time-frequency plane. Instead, we examined the visibility plane (referred to as the {\it uv}-plane in radio interferometry) to look for RFI-affected data. Since the sky brightness distribution is smooth, the visibility distribution is expected to be smooth as well. This makes it easier to identify outlier visibilities in the {\it uv}-plane for solar observations. These outliers were manually identified and flagged. Additionally, time or frequency slices with more than 80\% flagged data were flagged completely.

\subsection{Interferometric Imaging}\label{subsec:imaging}

Due to the much better RFI conditions at the MWA, we first examined the dynamic spectrum from MWA to identify the nature of solar emission features in the dynamic spectrum. The MWA dynamic spectrum is shown in the left of Figure \ref{fig:mwads}. For comparison, the uGMRT dynamic spectrum is also shown in the right panel of the same figure. MWA DS is flux density calibrated and has been produced using the algorithm described by \citet{oberoi2017}. The bottom panels of Figure \ref{fig:mwads} show the band-averaged flux density timeseries. Both uGMRT and MWA DS and timeseries match well in spectro-temporal morphology. It is evident that the DS shows numerous narrowband short duration bursts, reminiscent of type-I bursts and also show a high background flux density of approximately 30 SFU. In contrast, during a rather quiet condition, \citet{sharma2020} estimated the solar flux density with the MWA at similar frequencies to be only about 15-17 SFU, almost half the value observed here. Based on these reasons, we have identified the excess broadband persistent emission as the noise storm continuum and the narrowband short-lived emissions as the type-I bursts.

To facilitate data analysis, the entire frequency range was divided into four frequency chunks, 221--228 MHz, 229--235 MHz, 236--243 MHz, and 244--251 MHz, each approximately 6--7 MHz wide. The frequency width was chosen to match the typical bandwidth of type-I bursts in this frequency range \citep[e.g.][]{elgaroy1977, suresh2017, mohan2019b}. For each frequency chunk, we produced images at multiple snapshots corresponding to the peaks in the lightcurve (refer Figure \ref{fig:lightcurve_0_35}). Although this study focuses on type-I bursts, some time spans, when these bursts were absent, were also imaged to investigate the morphological differences, if any, between the type-I burst source(s) and the broadband continuum emission arising in the vicinity. The integration time was varied between images from 1.3 to 14 seconds, and was chosen to prevent any overlap between burst and nonburst times. Imaging was done using the CASA task \textsf{tclean}, using baselines longer than 2000$\lambda$. Avoiding the use of short baselines reduces correlated RFI contamination and hence provides better image quality. Additionally, \textit{uniform} weighting was used to maximize angular resolution and improve the quality of the instrumental point spread function (PSF). We refer to the PSF thus obtained as the synthesized PSF. 

\subsection{Image alignment}
\begin{figure*}
    \includegraphics[trim={1.3cm 0cm 3.2cm 1cm},clip,scale=0.735]{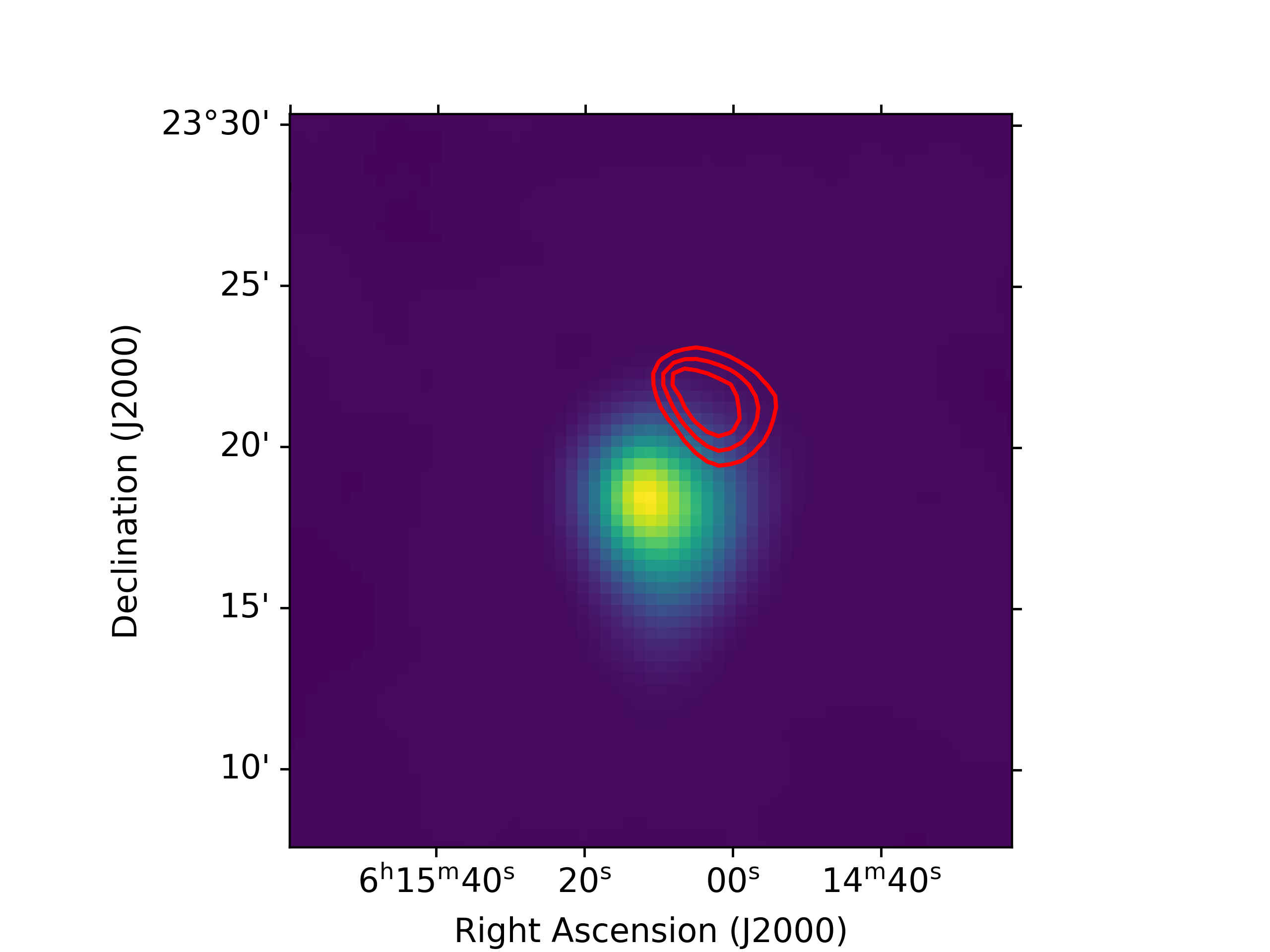}
    \includegraphics[trim={1cm 0cm 3.1cm 1.4cm},clip,scale=0.73]{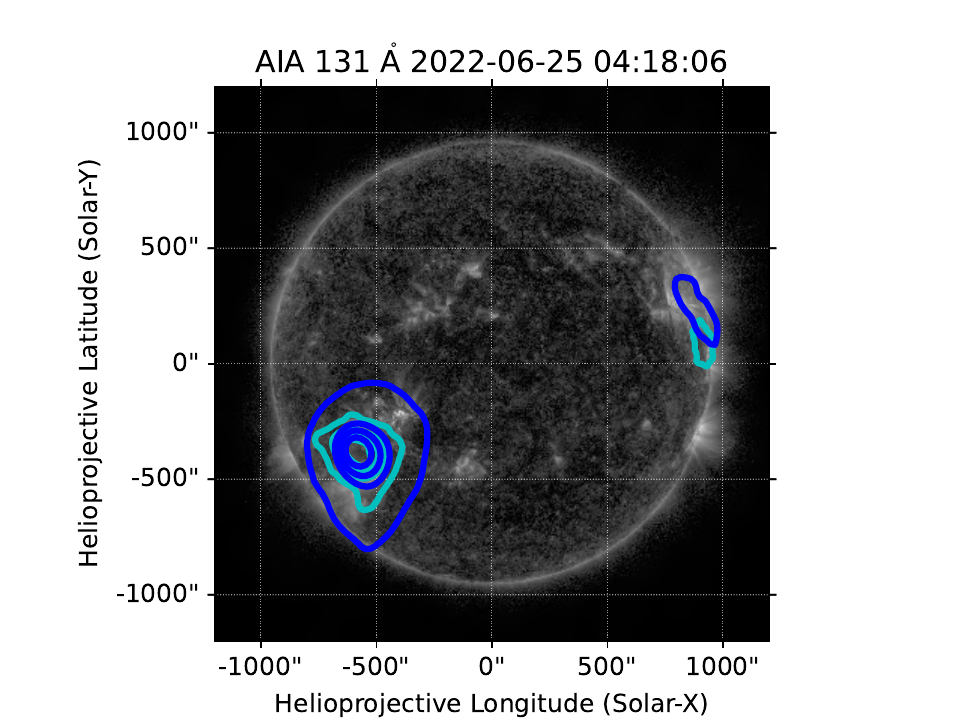}
    \caption{Left panel: Contours of the uGMRT image are overlaid on the MWA image. Both images correspond to 04:14:14 UT during a weak type-III radio burst. Right panel: Contours of the uGMRT and MWA images are overlaid on an AIA 131$\,$ image. The contours of the uGMRT and MWA images are cyan and blue respectively.}
    \label{fig:image_alignment}
\end{figure*}

Ionospheric refraction often shifts a source from its true location at low radio frequencies \citep[e.g.][]{cohen2009, loi2015}. Hence to obtain the true absolute location of a source, this refractive shift needs to be corrected. As the ionosphere over the two telescopes separated by large distances is independent, the shifts seen at each of them are usually different. For most astronomical sources, a comparison with the previously catalog positions is used to identify these shifts and correct for them, if needed. For a non-sidereal extended source like the Sun with a highly time and frequency variable morphology, which does not offer well-defined anchor points, correcting for ionospheric shift is more complicated. For MWA solar images, this shift is estimated by forcing the center of the quiet radio sun to lie at the center of optical solar disc. This process is described in detail in \citet{kansabanik_paircars1}. The uGMRT has a much smaller number of short baselines suitable for accurate imaging of the large angular scale quiet sun disc, and additionally, they are often corrupted by RFI.
Hence a direct application of this method is not possible for uGMRT data. However, since the simultaneous MWA images aligned with the true solar center are already available at overlapping frequencies, we simply align the solar emission in uGMRT images to that seen in the MWA images. 
Such a procedure will work best for simple source morphologies.
Fortunately, a weak type-III radio burst, which are characterized by compact emission morphologies, took place during these coordinated observations at 04:14:14 UT. 
In the left panel of Figure \ref{fig:image_alignment}, we overlay the uGMRT image contours over the MWA image prior to image alignment. The images are in the equatorial coordinate system. 
The significant offset in the emissions seen at the two instruments is evident, even though they correspond to the same physical source at the same time and frequency.
This shift is correctly by simply translating the uGMRT image to align the peaks of the observed source with that in the MWA image for the same spectral and temporal slice. It is clear that after this translation, as expected, the type III source in the GMRT image aligns very well with that of the MWA image. However the faint extended emission seen on the western limb, is not aligned for the 2 images. The offset is much smaller than that observed for the type III before the correction. Such offsets are not unexpected because of the differential refraction of the ionosphere \citep{cohen2009}. GMRT has much longer baselines, as compared to the MWA and hence is sensitive to small scale ionospheric density inhomogeneities. These inhomogeneities can vary over the $32^{'}$ solar disc, particularly during the noon time \citep{cohen2009}. We also see that the extent of the western limb source is much larger in the MWA image, than the source in the GMRT image. This is because of the different baselines used for producing the images. For the GMRT image, we have only used baselines above 2k$\lambda$, which is only sensitive to structures below $80\arcsec$. No such limitation exists for the MWA image. 

This work primarily depends on the morphology of the radio sources and not very sensitive to their absolute astrometric accuracy. Hence although the ionospheric shift is time variable, we have applied the offset using the method described above to all other time and frequency slices used here.
After this translation, the uGMRT images are then transformed into the helioprojective coordinate system. The final radio images after these transformations are shown by blue and cyan contours for the MWA and uGMRT, respectively, in the right panel of Figure \ref{fig:image_alignment}, overlaid on a nearby AIA 131\AA image. The alignment between the MWA and uGMRT images is self-evident.

\section{Results}\label{sec:results}
\begin{figure}
    \centering
    \includegraphics[trim={1.3cm 0cm 2cm 0.5cm},clip,scale=0.68]{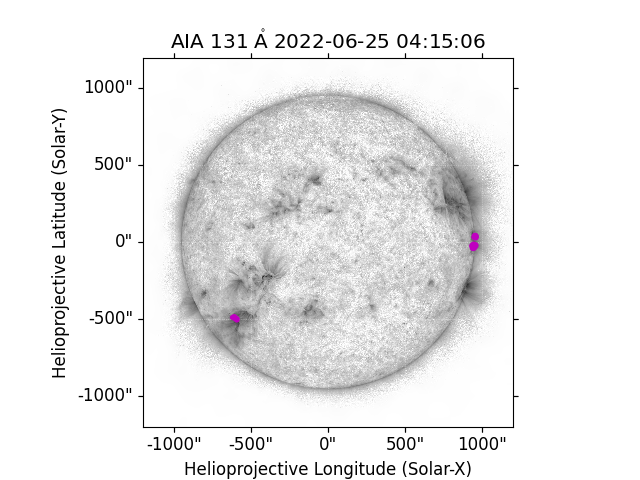}
    \caption{Contours of an example radio image of the Sun for the 244--251 MHz band at 04:14:30 is overlaid on a nearby AIA 131\AA$\,$ image. The AIA image has an inverted colorscale, where darker shades shows brighter regions. The contour levels are at -0.4, 0.4, 0.6, and 0.8 times the peak of the radio image.  Solid and dashed lines represent positive and negative contours, respectively.}
    \label{fig:aia_overlay_full_sun}
\end{figure}

Figure \ref{fig:lightcurve_0_35} shows the cross-correlation amplitude in the 244--251 MHz band as an example. Each line represents a timeseries averaged over consecutive 50 baselines, arranged in ascending order by baseline length at the mentioned frequency range. Variations in the cross-correlation amplitude are directly related to variability in the source properties and are often used as a proxy for lightcurve of a source, especially for compact sources. The vertical bars mark the burst times which have been imaged. Figure \ref{fig:aia_overlay_full_sun} shows radio contours overlaid on an 131\AA$\,$ image from the Atmospheric Imaging Assembly \citep[AIA,][]{lemen2012} onboard the Solar Dynamics Observatory \citep[SDO,][]{pesnell2012}. The contours are at -0.4, 0.4, 0.6, and 0.8 times the peak. Solid and dashed lines represent positive and negative contours, respectively. The angular resolution of this image and all other images at this frequency range is $17.5\arcsec \times 9.2\arcsec$. PSF size for images at the other three frequency bands increases with decrease in frequency. All radio images, including this one, do not show any evidence of extended emission from the solar disc, as only baselines longer than $2000\lambda$ were used for imaging. This implies that structures with angular scales $\gtrsim80\arcsec$ will not be captured in these radio images. We observe that there are two clusters of compact sources, one located close to the eastern limb and the other located close to the western limb of the Sun. This paper focuses on the brighter of the two, located at the western limb. This source is present over the entire duration of the observation and has a bandwidth exceeding 30 MHz. Combining this with the fact that this is also located close to an active region, we regard this source to be a type-I noise storm.

\begin{figure*}
    \centering
    \includegraphics[trim={0.0cm 0cm 0cm 0.0cm},clip,scale=0.43]{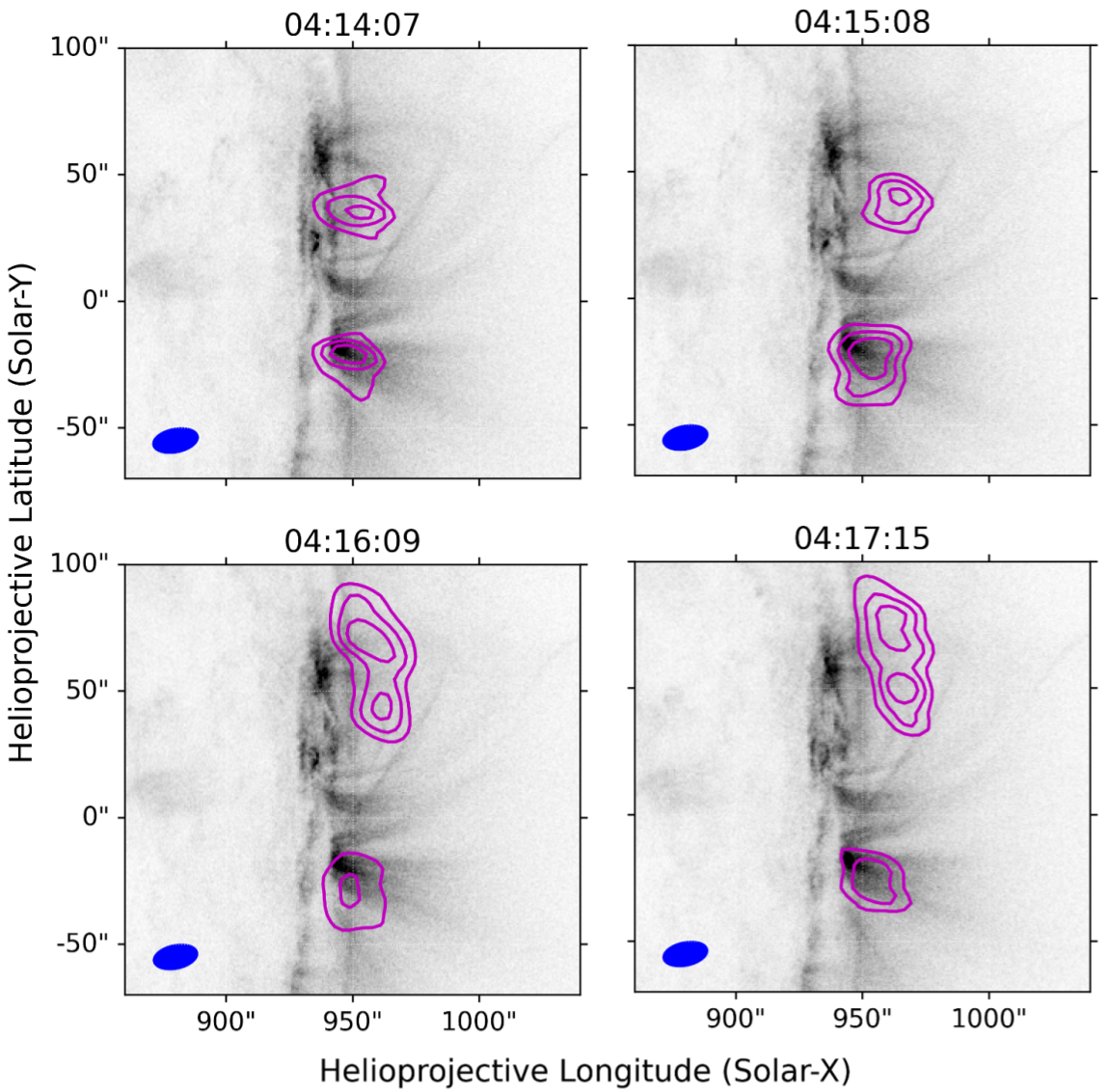}
    \caption{Four example radio images during type-I bursts from 244--251 MHz band are shown with magenta contours, overlaid over a nearby AIA 131\AA$\,$ image. The colorscale of the AIA image is inverted greyscale. The title of each panel shows the time of the radio image. The contour levels are at -0.4, 0.4, 0.6, and 0.8 times the peak of the individual radio images. The blue ellipse at the bottom left corner shows the resolution of the radio image.}
    \label{fig:zoomed_western_source_peak_times}
\end{figure*}

\subsection{Temporal Variations in Source Morphology} \label{subsec:temporal_variation}
Figure \ref{fig:zoomed_western_source_peak_times} shows some example solar radio maps zoomed in at the western limb source. These are chosen from the time spans marked by red boxes in Figure \ref{fig:lightcurve_0_35}. The integration time used for each image is determined primarily by the duration of the emission feature and marked by the red vertical bars. Contours for each radio image are overlaid on a nearby 131\AA$\,$ image from the AIA. The blue filled-ellipse at the bottom left corner of each image shows the synthesized PSF of the array. We find that for each of the four times shown, more than one source is observed. Two well-resolved components which are separated by about an arcminute are clearly seen. Additionally, while the southern source always has a relatively simple morphology, the northern source shows the presence of an additional component after approximately 04:15:53 UTC.

\begin{figure}
    \centering
    \includegraphics[trim={2cm 0cm 2cm 0.5cm},clip,scale=0.55]{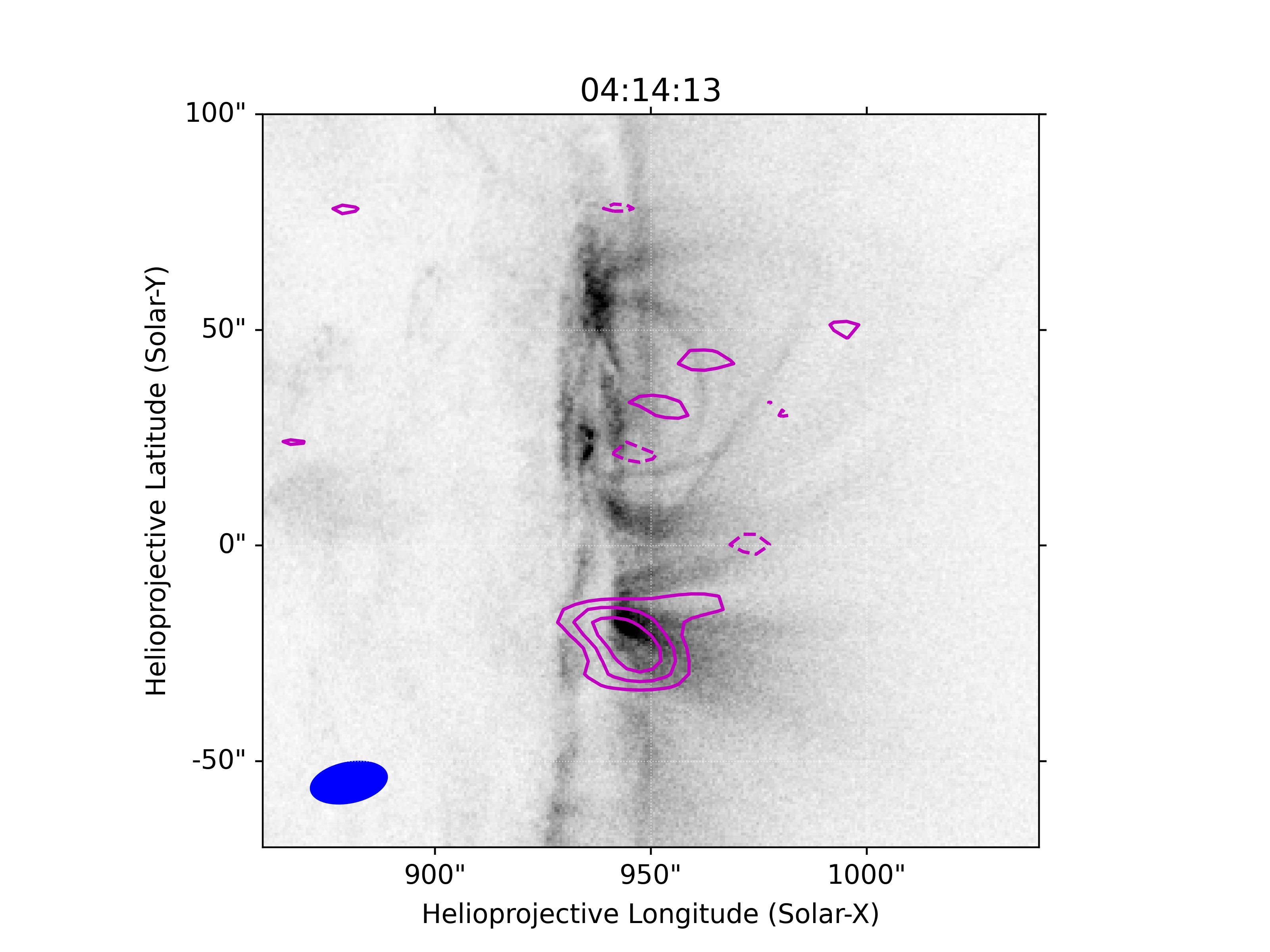}
    \caption{Contours from a radio image made at 244--251 MHz and 04:14:13--04:14:23, have been overlaid on a corresponding nearby AIA 131\AA $\,$ image. During this time, no type-I burst was detected. The contour levels are at -0.4, 0.4, 0.6, and 0.8 times the peak of the radio image. The blue ellipse at the bottom left corner shows the resolution of the radio image.}
    \label{fig:no_enhancement_spw_0_35}
\end{figure}

We believe that these structures are similar to the knots reported by \citet{mercier2015}. When the solar flux density fall to lower values, corresponding to timestamps outside the gray boxes marked in Figure \ref{fig:lightcurve_0_35}, flux densities of both sources reduce significantly. However, the amount by which they decrease is not the same for the northern and the southern sources. An extreme example of this is shown in Figure \ref{fig:no_enhancement_spw_0_35}, where the northern sources fall below the detection threshold, while the southern source is detected with good significance. This image was generated by averaging 04:14:13 --04:14:23 UTC. For comparison, the top left panel of Figure \ref{fig:zoomed_western_source_peak_times} corresponds to 04:14:12 -- 04:14:13 UTC, where we see a type-I burst. During this period, the northern source was comparable to the southern source. However, in the immediately adjacent 10-second period, when the burst emission is absent, the northern source has faded below the detection threshold.

\subsection{Spectral Variability} \label{subsec:spectral_variation}

\begin{figure*}
    \centering
    \includegraphics[trim={0 0 0cm 0},clip,scale=0.205]{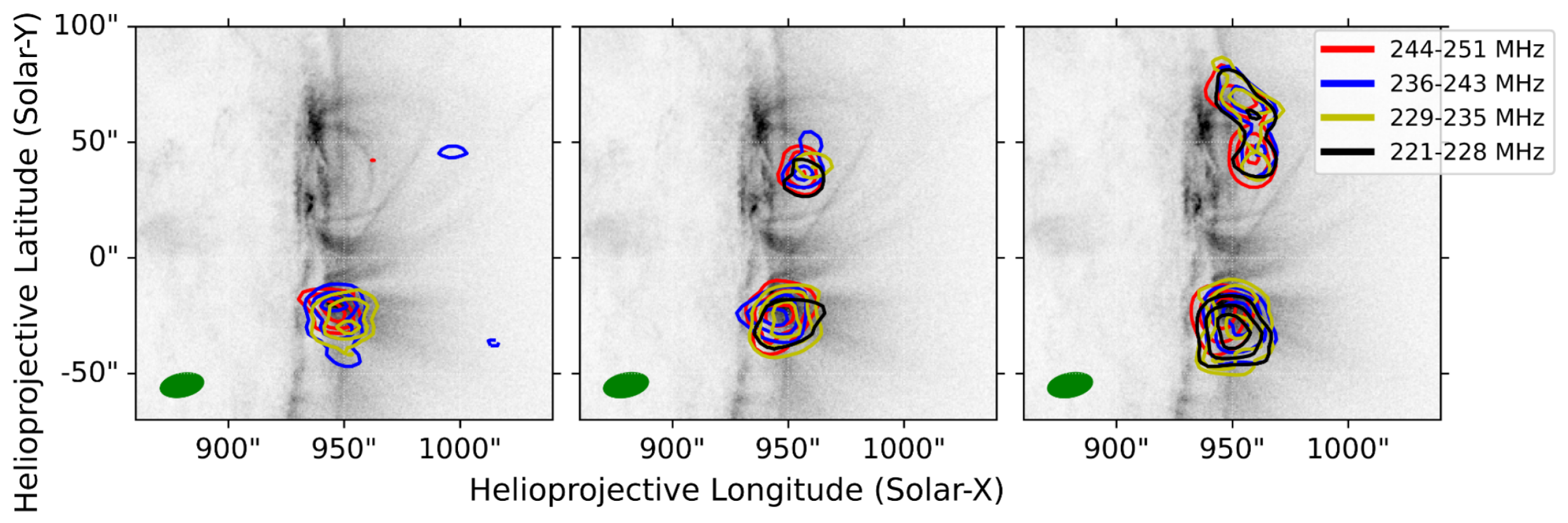}
    \caption{Radio images from different frequency bands at similar times. Left panel: Contours from 244--251 MHz, 236--243 MHz, and 229--235 MHz are shown. The corresponding image times are 04:14:13--04:14:23, 04:14:13--04:14:16 and 04:14:13--04:14:16 respectively. During these times no type-I burst is detected at the corresponding frequencies. Middle panel: Contours from 244--251 MHz, 236--243 MHz, 229--235 MHz, and 221--228 MHz has been shown. The corresponding image times are 04:14:28--04:14:30, 04:14:28--04:14:30, 04:14:28--04:14:30 and 04:14:06--04:14:07 respectively. These times correspond to type-I bursts at their corresponding frequencies. Right panel: Contours from 244--251 MHz, 236--243 MHz, 229--235 MHz, and 221--228 MHz are shown. The corresponding image times are 04:15:52--04:15:57, 04:16:02--04:16:04, 04:16:03--04:16:04 and 04:16:03--04:16:04 respectively. They correspond to the times when a morphology change is seen in the northern source at each of the four frequency bands. The contour levels of the image at 04:14:01--04:14:02 are at -0.3, 0.3, 0.6 and 0.9 times the peak. Contour levels for all other images are -0.5,0.5, 0.7, and 0.9 times the peak. The green ellipses at the bottom left corner show the resolution of the radio images.}
    \label{fig:mult_freq_source_morphology}
\end{figure*}
 
Similar to other previous works, we also find that type-I bursts are narrowband in nature. These data show several instances where type-I burst is present in one frequency band and not in the other three. However, the strongest burst, which occurred around 04:16:08, was detected across the entire frequency band. We note that in multiple instances at the lower two frequency bands, the western limb source was not detected due to dynamic range limitations arising from the presence of the bright eastern limb source.
 
Figure \ref{fig:mult_freq_source_morphology} shows the source morphology at some example times of interest at the frequencies bands studied here. The left panel shows the source morphology around 04:14:13, when the type-I burst is not detected in any of the frequency bands. For this time, no emission was detected at the lowest frequency band. The middle and right panels show the morphology around 04:14:28 and 04:15:52 respectively. For each frequency, we have shown AIA images nearby the stated time where we observe a type-I burst. It is evident that the source morphology is quite similar at all the frequency bands. Additionally, as noted in Section \ref{subsec:temporal_variation}, the morphology of the northern source changes around 04:15:48, and Figure \ref{fig:mult_freq_source_morphology} shows that this change is observed at the other three frequency bands as well at similar times. 

\subsection{Source Structure at Small Spatial Scales} 
\begin{figure*}
     \centering
     \includegraphics[trim={1.5cm 0cm 2.9cm 1.3cm},clip,scale=0.73]{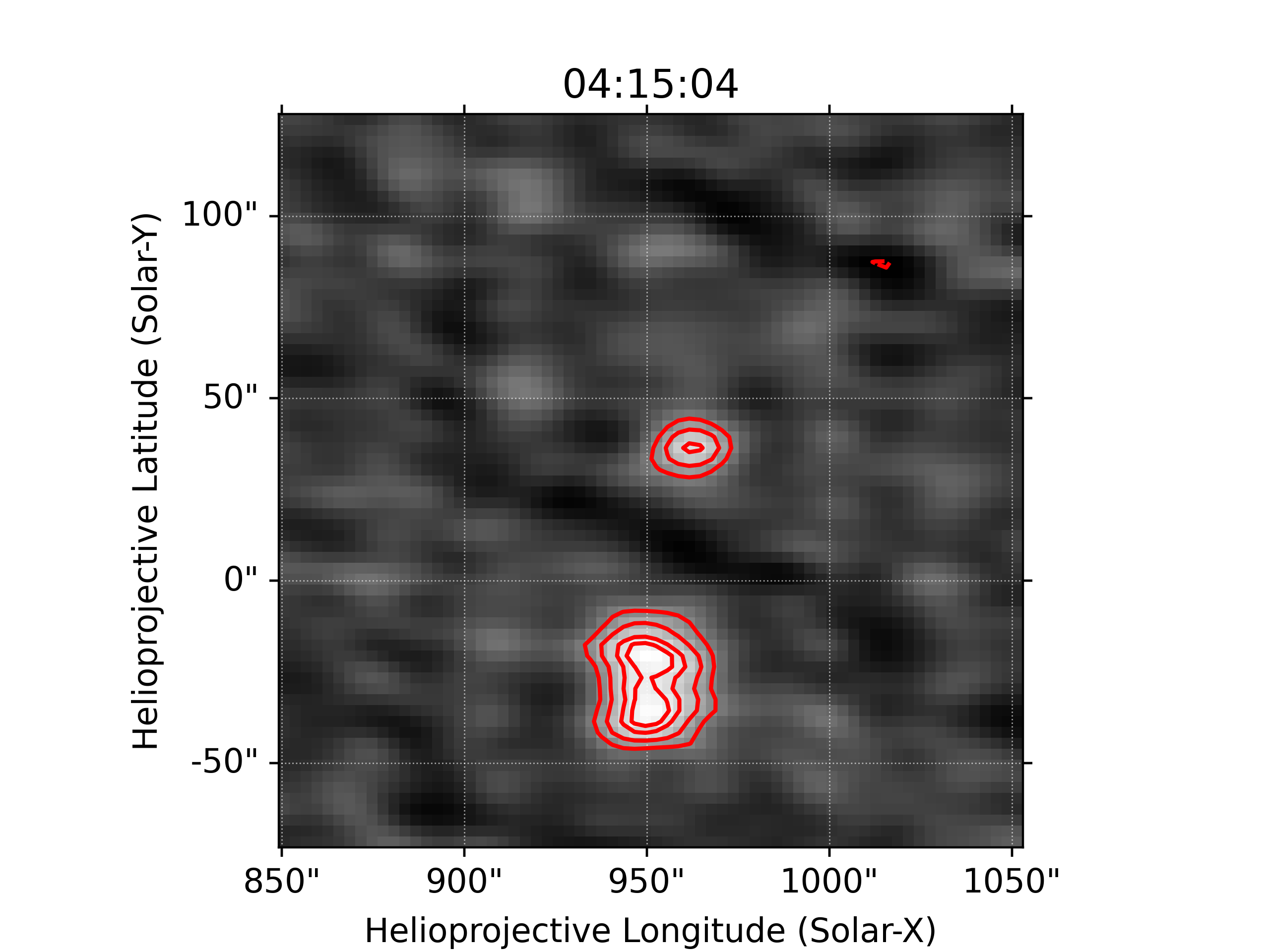}
     \includegraphics[trim={2.1cm 0cm 2.9cm 1.3cm},clip,scale=0.73]{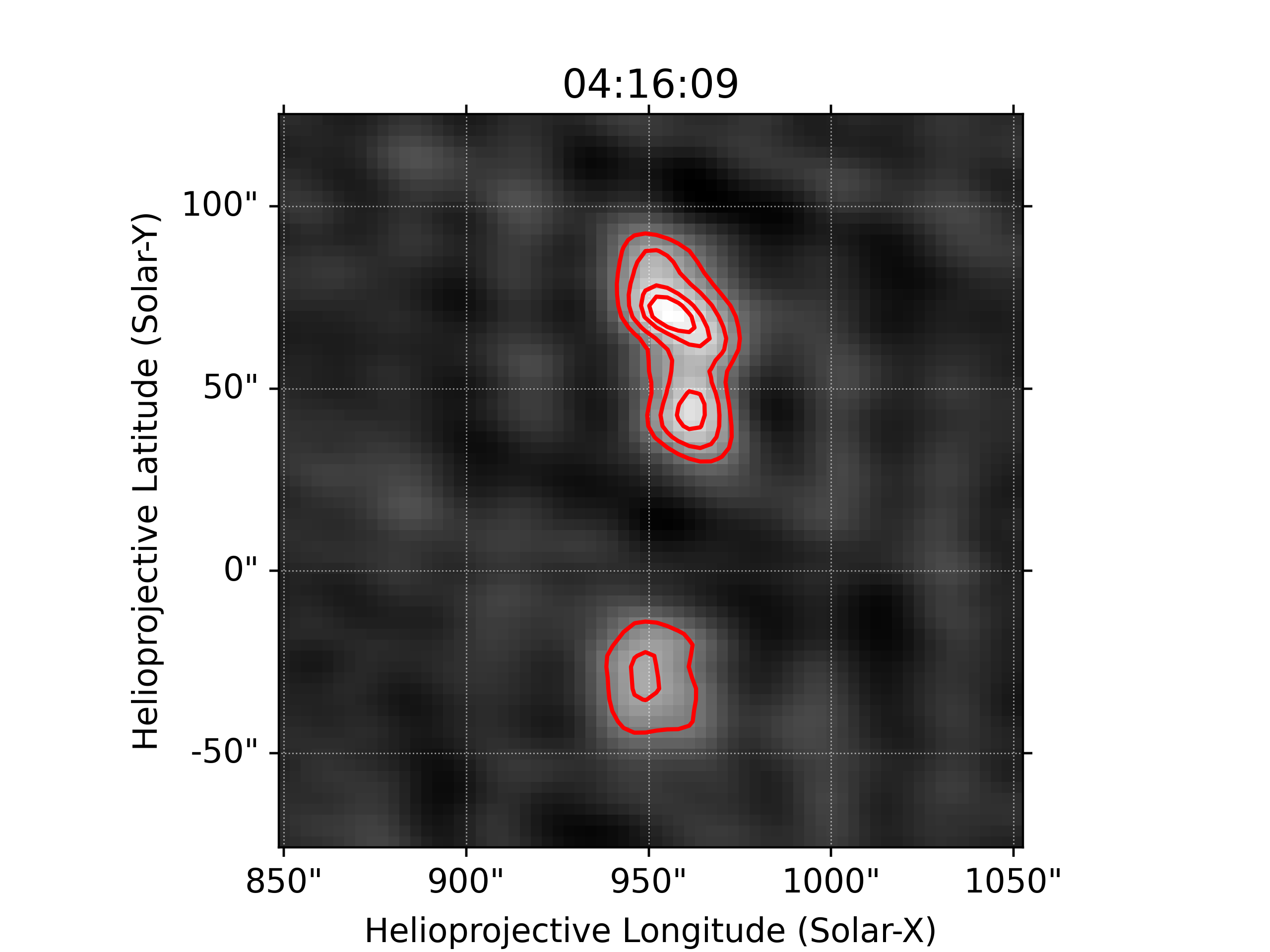}
     \caption{Two solar radio images showing some of the smallest source sizes measured. The time in the title corresponds to that of the radio image.}
     \label{fig:smallest_scale}
 \end{figure*}

A key motivation of this work was to investigate the smallest spatial scales visible at these frequencies. Past works on noise storms have shown the smallest angular sizes to be about $30\arcsec$--$35\arcsec$ \citep{mercier2015}. Figure \ref{fig:smallest_scale} shows two example images in the 244--251 MHz band the frequency at which some of the smallest source structures are visible. To determine the source size, we fitted the source of interest with Gaussian(s) and a constant offset. For the left and right panels of the same figure, a single and double Gaussian model, respectively, was used to fit the northern source. The northern source in the left panel of this figure is well described by a Gaussian of size $18(2)\arcsec \times 13(1.7)\arcsec$, on top of a constant flux density of 3 Jy/beam. The numbers within the brackets are the uncertainties on the corresponding parameters. In a radio image, the observed source size is actually a convolution of the ``true" size of the source with the instrumental PSF, which must be deconvolved to determine the ``true" source size. Since the observed Gaussian size is very close to the instrumental PSF ($17.5\arcsec\times 9.2\arcsec$), the source is essentially unresolved. We also find that the integrated and peak flux density of the source obtained from the Gaussian fit are $36\pm7$ Jy and $24 \pm 3$ Jy/beam respectively. We find that after subtracting off the baseline flux, the integrated and peak flux densities are consistent within errors. This corroborates our original conclusion that the source is indeed unresolved. The maximum source size is then given by the instrumental beam. However, the true source size can be significantly smaller. 
For the right panel of the same figure, the best fit Gaussian components are $38\arcsec\times 23\arcsec$ and $24\arcsec\times 18\arcsec$ for sources centered at ($956\arcsec,71\arcsec$) and ($961\arcsec, 45\arcsec$), respectively. The deconvolved source sizes of these respective sources are $36(2)\arcsec\times 18(2)\arcsec$ and $20(2)\arcsec \times 10(4)\arcsec$. Similar source sizes are seen for other times and frequencies and it is evident that even prior to deconvolving the effect of the instrumental beam, 
many of the observed source sizes are already significantly smaller than the smallest sizes reported earlier. 
 \begin{figure}
     \centering
     \includegraphics[trim={0.3cm 0cm 1.2cm 0cm},clip,scale=0.55]{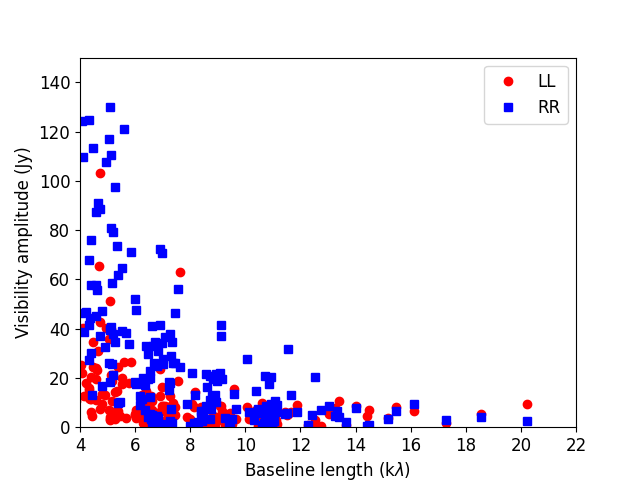}
     \caption{Visibility amplitude for 244-251 MHz band has been plotted as a function of the baseline length, where the baseline length has been measured in units of $k\lambda$. All times and frequencies between 244-251 MHz have been averaged for this figure.
     }
     \label{fig:uvwave_plot}
 \end{figure}

The presence of small spatial scales can also be inferred by examining the observed visibilities as a function of baseline length. Figure \ref{fig:uvwave_plot} shows the variation of visibility amplitudes, averaged over 7 MHz and the entire time duration, with the baseline length. We have only shown baselines whose lengths are greater than 4000$\lambda$. The blue and red symbols show the averaged visibilities for RR and LL polarizations, respectively. We find significant power at baseline lengths greater than 5000$\lambda$, corresponding to an angular scale of approximately $40\arcsec$. This is also the length scale beyond which most of the past studies did not detect any small-scale structure \citep{zlobec1992,lang1987}. To investigate if the visibility amplitudes seen at these and even longer baseline lengths are reliable, we have estimated the noise on individual visibility measurements. These solar observations are in the strong noise regime (also known as self-noise), implying that the observed noise is essentially independent of the instrumental characteristics and the error on each visibility is given by $S/\sqrt{\Delta \nu \Delta t}$, where $S$ is the source flux density seen by the instrument and $\Delta t$ and $\Delta \nu$ are the time and frequency spans over which the visibility is measured \citep{Radhakrishnan1999,morgan2021}. Using the band-averaged lightcurve from the flux-density calibrated MWA dynamic spectrum shown in Figure \ref{fig:mwads}, we estimate that the mean flux-density during the observation is $\sim$30 SFU (1 SFU (solar flux unit) = $10^4$ Jy). This translates to a theoretical uncertainty of 6 Jy on each visibility for the time-frequency integration used here. Thus, it is evident from Figure \ref{fig:uvwave_plot} that many visibilities from baselines with lengths $\lesssim10000\lambda$ have a signal-to-noise ratio $\gtrsim 5$. 
This implies that even on individual baselines, we detect emission at angular scales about half the size detected in most previous works. 
We note that strictly speaking, $S$ is the apparent flux seen by the baseline for which the source noise contribution is being estimated. The flux density of 30 SFU used to estimate this uncertainty is the total flux density of the Sun (disc integrated), almost all of which is resolved out at baselines $\gtrsim 2k\lambda$. This is also the reason why we do not see any extended emission in the images shown here. Hence, the theoretical uncertainty estimated here is an extreme upper limit, it assumes that the entire flux density of the Sun remains unresolved even at the longest baselines. Even in this extreme situation, we find evidence for structures smaller than $20\arcsec$ from individual visibilities themselves, independent of any imaging. Estimating the true source noise on these long baselines requires determining the flux at the corresponding angular scale. This necessarily involves imaging, and the existence of small-scale structures based on imaging studies has already been discussed earlier in this section.

\begin{figure}[htbp]
    \centering
    \includegraphics[trim={0.3cm 0 1cm 0.6cm},clip,scale=0.58]{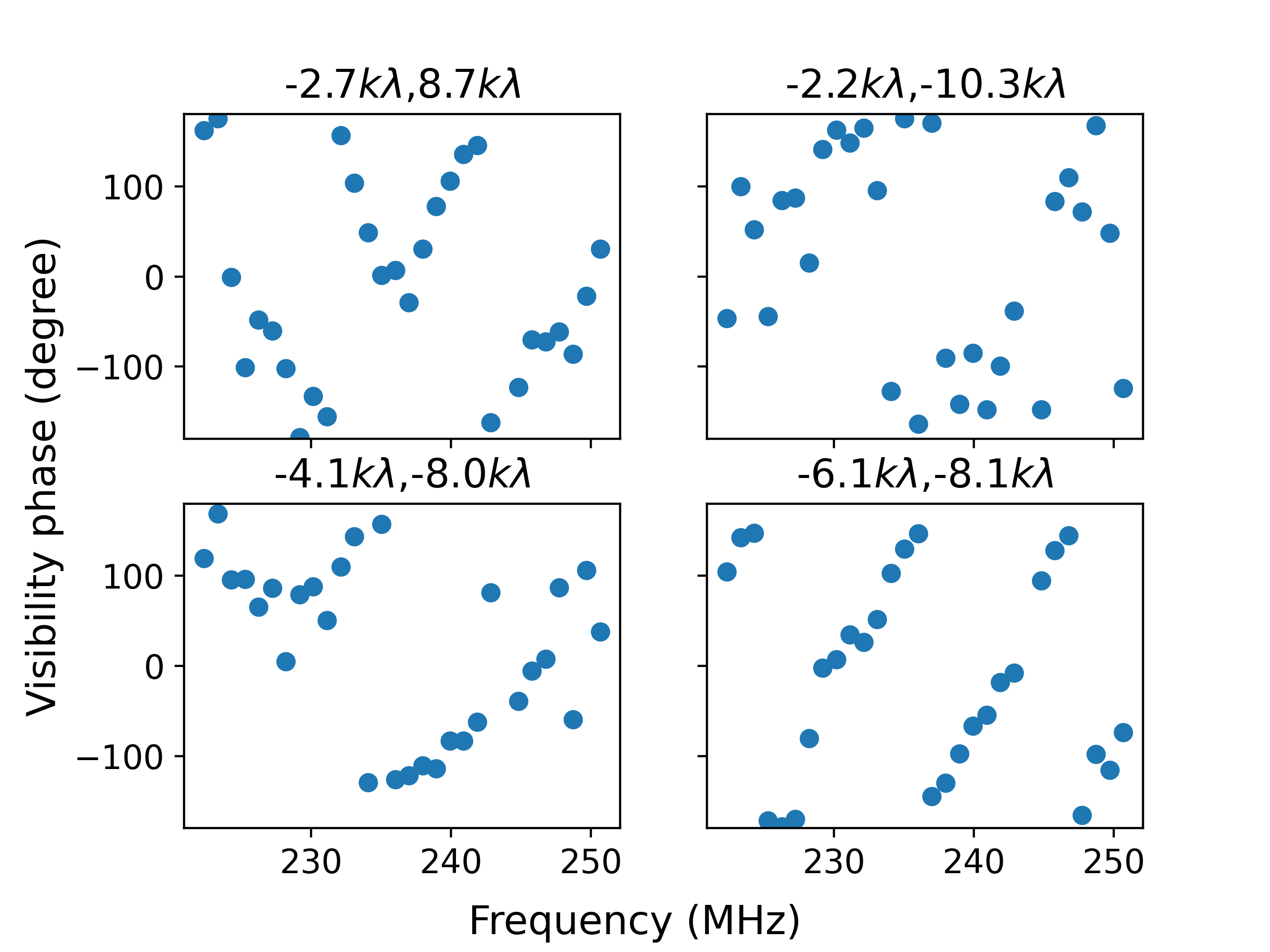}
    \caption{This figure shows the phase variation as a function of frequency for a few chosen baselines. The title shows the median $(u,v)$. }
    \label{fig:phase_coherent}
\end{figure}

However, the condition that significant power on long baselines is sufficient evidence for the presence of structure at that length scale can potentially break down in the presence of multiple sources. The contribution of each source to a baseline depends on the relative orientation of the baseline with respect to the source, as well as the angular scales at which the source emits and the one probed by the baseline. The total power seen by a baseline is hence the superposition of contributions from each of the individual sources to that specific baseline. 
Hence there is a possibility, however small, that contributions from different sources can interfere constructively to produce detectable power in some long baselines, even if none of the sources themselves produce emission at angular scales to which that long baseline corresponds. 
To investigate whether this could be true in the present case, as shown in Figure \ref{fig:phase_coherent}, we examine the phase variation as a function of frequency for a few selected baselines. All the baselines chosen have lengths between $7$--$12k\lambda$. The projected length of each baseline changes with time due to the rotation of the Earth. The title of each panel of Figure \ref{fig:phase_coherent} shows the median coordinates $(u,v)$ in the Fourier plane ({\it uv-}plane), where the median has been taken over frequency and time. We have averaged over 0.98 MHz to detect the signal over noise while also preserving the frequency structure. 
Systematic and well-defined spectral trends are seen for all of the baselines, and the median lengths lie the range from $8.9$--$10.5k\lambda$.
If the amplitude peaks seen in the longer baselines in Figure \ref{fig:uvwave_plot} are simply `noise' peaks due to chance constructive interferences, they are not expected to give rise to the
observed systematic phase variation with frequency for multiple baselines populating different part of the $uv$ plane.
This provides robust confirmation for the presence for sources at small angular scales during type I noise storms.

\section{Discussion} \label{sec:discussion}

\subsection{Nature of type-I bursts}

Multiple earlier works have revealed that type-I bursts have bandwidths of about 6--7 MHz. \citet{mondal2021} specifically searched for the correlation between bursts at well-separated frequencies and did not find any.  \citet{mohan2019b} suggested that the weak reconnections happening at the noise storm sites produce very weak electron beams which quickly get collisionally damped, leading to their small bandwidths. Hence it can be concluded that electron beams responsible for producing type-I bursts at widely separated frequencies are, in general, different, particularly for weak bursts. This also suggests that structures inside the noise storm region likely correspond to regions of magnetic reconnection and subsequent electron acceleration. We also find that while during the bursts, both the northern and southern sources get brighter, the magnitude of this enhancement can be significantly different for the two sources.

This work also reveals an interesting correlation between the reconnections happening at different coronal heights, which accelerates electrons and ultimately leads to plasma emission at different frequencies. We find that the morphology of the type-I burst sources is very similar across the 30 MHz observing band studied here. Additionally, changes in the source morphology also occurred across the entire observing band at very similar times as well. This can be explained if we assume that whatever source is causing the magnetic reconnections to happen at different heights is actually traveling along the coronal loops. When magnetic reconnection starts at new locations, due to some disturbance/wave, that disturbance/wave travels along the loop and induces magnetic reconnection at different heights as well. This phenomenon will show up as new type-I sources appearing at all frequencies at similar times. The exact time of appearance however, will depend on the disturbance, its propagation speed and as well as the details of the electron acceleration and how it gets converted to plasma emission. \citet{mohan2019b} and \citet{mondal2021} found evidence of sausage mode waves inducing reconnection at the type-I sites, and hence can serve the role of a traveling disturbance required to explain these observations. However, we emphasize that investigating the detailed nature of the driver, as required to interpret these observations, is beyond the scope of this work.

\subsection{Source size at the smallest scales}

In this work, we have presented evidence for sources with angular sizes as low as $\lesssim 9\arcsec$, about a third of the smallest size reported earlier.
Multiple studies have been devoted to quantifying the impact of scattering on the apparent observed size of a point source. Their outcome, naturally, is highly dependent on the details of the model used to describe the coronal inhomogenities and prescription used for modeling scattering.
For instance, while the formalisms by \citet{bastian1994} and \citet{kontar2019}, both predict source sizes much larger than shown here, that by \citet{subramanian2011} predicts that the minimum possible source size can be smaller than seen here.
While the objective of this work is limited to presenting observations of unexpectedly compact coronal sources, the following discussion is helpful in understanding the origin of large discrepancies in predictions from different formalisms and models, and also between different observations:
\begin{enumerate}
    \item {\bf Spherical vs plane wave propagation} -- \citet{subramanian2011} showed that for an identical scattering medium, spherical wave propagation model predicts a source size about two orders of magnitude smaller than that predicted using the plane wave approximation.
   
    \item  {\bf Choice of inner scale} -- Both \citet{subramanian2011} and \citet{kontar2019} noted the importance of the choice of inner scales of turbulence used. However, the values used in these works are very different and may have a role in these widely varying predictions. 
     
    \item {\bf Normalization of structure functions} -- Another place where these models differ is their choice of normalizations for structure functions. The level of density fluctuations close to the source is extremely important for determining the expected minimum source size, and hence differences in choice of this crucial number can also lead to very different predictions.
    
    \item {\bf Anistopic scattering and viewing angle} -- \citet{kontar2019} studied the effect of anisotropic scattering and viewing angle. They found that a source can look small along one of the spatial dimensions if it is located away from the disc center. Interestingly, in all the works that report small source size \citep[e.g.][]{mercier2006, mercier2015, mugundhan2018}, and this work, the source of interest was located close to the limb. In fact, \citet{mercier2015} also noted that no small-scale structure is seen in the noise storm sources located close to the disc center and they were much more extended as compared to the one located close to the limb. \citet{zlobec1992} did not detect any small-scale structure and for their observations, the noise storm was located close to the disc center as well. 
    In the observations available thus far, sources with small angular sizes are found to predominantly lie close to the limb.
    This is surprising because emission from the limb source has a longer propagation path through the corona to an Earth-bound observer, compared to a source on the disc center. Hence a limb source is expected to be subject to more severe propagation effects like angular broadening due to scattering. It is, however, possible that this apparent contradiction, of detection of smaller angular scale structures from limb sources, might arise simply due to the very small sample size. Hence this needs to be investigated more systematically using a bigger sample.

    \item {\bf Dynamic nature of the solar corona} -- It is well accepted that the solar corona is a dynamic and highly turbulent medium. Hence it is not hard to imagine that in some instances the properties of the medium are such that it becomes possible to observe very small scale features, than possible otherwise. 
\end{enumerate}

We also note here that the source size observed for type-III solar radio bursts at similar frequencies has always been found to be much larger than seen here, and type-I sources in general at similar frequencies. 
Since scattering effects do not distinguish between the emission source, it is hard to explain this consistent difference based on the properties of the scattering medium alone unless the propagating medium surrounding the type-I source is different from that of a type-III source. 
We believe that understanding this difference between the extremely small source size of type-I noise storms can also shed more light on the emission sites of type-I bursts and the surrounding environment of the type-I sources as well. 

\section{Conclusion} \label{sec:conclusion}
This work presents the first uGMRT study of the solar corona. We have used the high angular resolution data from the uGMRT to resolve type-I noise storms and have detected structures of angular scales $\sim 9\arcsec-20\arcsec$, about 2--3 times smaller than the smallest structures reported to date.
Our study shows that while the individual type-I bursts are themselves narrowband in nature, the morphology of the type-I bursts is very similar across the much wider frequency range covered by our observations. 
During our observations, the morphology of one of the burst sources changes from a single compact but resolved source to two adjacent resolved sources.
Additionally, the change in the morphology of the bursts happens around the same time across the entire band of observations. 
We interpret this as arising due to a traveling disturbance inducing or triggering independent reconnections at different heights as it traverses a magnetic bundle, and leads to the generation of localized electron beams that give rise to the observed narrowband type-I bursts. 
This scenario allows for the morphology of emission to change in tandem across a wide range of frequencies, while individual type-I bursts remain narrowband.
This suggests that a new traveling disturbance originated around 04:15:47 UTC, and as it traversed the range of coronal heights covered by our observation bandwidth, it led to the increased complexity in the morphology of the northern source. Thus these high spatial resolution observations suggest a natural mechanism to simultaneously explain both -- the narrowband nature of type-I bursts and their obvious relationship with the broadband noise storm emission. 

This study provides robust evidence for the presence of small angular scale structures in metrewave type-I noise storms and type-I bursts. High angular resolution imaging studies, similar to the one presented here, can serve multiple useful purposes -- on the one hand they provide an opportunity to constrain and characterize coronal scattering models 
; and on the other, they allow developing connections between the type-Is in low radio frequency emissions with those seen in higher energy bands like X-rays and EUV, which are necessary to build a better understanding of the underlying phenomena than possible with observations at any given waveband. 
The uGMRT, with its high angular resolution and sensitivity at low radio frequencies, is well poised to make a significant contribution to this effort.

\facilities{Murchison Widefield Array \citep[MWA;][]{Tingay2013,Wayth2018},Upgraded Giant Metrewave Radio Telescope \citep[uGMRT;][]{Swarup2000,gupta2017}, Solar Dynamics Observatory \citep[SDO;][]{pesnell2012}.}

\software{CASA \citep{CASA2022}, P-AIRCARS \citep{kansabanik_paircars_2,paircars_zenodo}, SunPy \citep{sunpy_community2020,sunpy2.1.0}, astropy \citep{astropy:2013,astropy:2018,astropy:2022}, matplotlib \citep{Hunter:2007}, Numpy \citep{Harris2020},Scipy \citep[][]{Scipy2020}, Python 3 \citep[][]{python3},}\\

\noindent This work uses observations from the Giant Metrewave Radio Telescope (GMRT) run by the National Centre for Radio Astrophysics of the Tata Institute of Fundamental Research. We thank the staff of the GMRT that made GMRT operational. We thank observers at GMRT who made these non-trivial solar observations possible. This scientific work makes use of the Murchison Radio-astronomy Observatory (MRO), operated by the Commonwealth Scientific and Industrial Research Organisation (CSIRO). We acknowledge the Wajarri Yamatji people as the traditional owners of the Observatory site. Support for the operation of the MWA is provided by the Australian Government's National Collaborative Research Infrastructure Strategy (NCRIS), under a contract to Curtin University administered by Astronomy Australia Limited. We acknowledge the Pawsey Supercomputing Centre, which is supported by the Western Australian and Australian Governments.  We also thank the anonymous referee for the comments and suggestions, which have helped improve the clarity and presentation of this work. S.M. acknowledges the partial support of the NASA ECIP grant \#80NSSC21K0623 to NJIT for the publication fee. D.K. acknowledges support for this research by the NASA Living with a Star Jack Eddy Postdoctoral Fellowship
Program, administered by UCAR’s Cooperative Programs for the Advancement of Earth System
Science (CPAESS) under award \#80NSSC22M0097. 
D.O. and S.D. acknowledge the support of the Department of Atomic Energy, Government of India, under project no. 12-R\&D-TFR-5.02-0700.

\bibliography{apj_version}{}
\bibliographystyle{aasjournal}

\end{document}